\documentclass[12pt]{article}
\usepackage{amsfonts,latexsym,fullpage,amsmath,latexsym,amssymb,epsf}

\date{August 1, 2002}

\newcommand{\cA}{{\cal A}}

\newcommand{\beq}{\begin{equation}}
\newcommand{\eeq}{\end{equation}}
\newcommand{\beqy}{\begin{eqnarray}}
\newcommand{\eeqy}{\end{eqnarray}}

\newenvironment{Definition*}{{\bf Definition}}{}

\def\C{{\mathbb{C}}}

\title{\Large \textbf{ 
Quantum noise influencing 
human behaviour  \\ could fake effectiveness of drugs in
clinical trials}} \author{ Dominik
Janzing\thanks{e-mail: 
janzing@ira.uka.de} \,\, and Thomas
Beth \\ \small Institut f{\"u}r Algorithmen und Kognitive Systeme,
Universit{\"a}t Karlsruhe,\\[-1ex] \small Am Fasanengarten 5,
D-76\,131 Karlsruhe, Germany}

\begin{document}

\maketitle

\abstract{
To test the effectiveness of a drug one can advice
two randomly selected groups of patients to take or not to take it,
respectively. It is  well-known 
that the causal effect cannot be identified 
if not all patients comply.
This holds even when the non-compliers can be identified afterwards
since latent factors like patient's personality can influence both his 
decision and his physical response.
However, one can still give bounds on the effectiveness 
of the drug depending on the rate of compliance.
Remarkably, the proofs of these bounds given in the literature 
rely on models that represent 
all relevant  latent factors (including noise) 
by hidden {\it classical} variables.
In strong analogy to the violation
of Bell's inequality, 
some of these bounds fail if
patient's behavior is influenced by latent quantum processes
(e.g.\ in his nervous system). Quantum effects could fake an 
increase of the recovery rate 
by about  $13\%$
although the drug  would hurt as many patients as it would help
if everyone took it.
The other bounds are true even in the quantum  case.

We do not present any realistic model showing this effect, we only point
out that the physics of decision making could be relevant for the causal
interpretation of every-day life statistical data. 
}

\section{The problem of noncompliance \\ in randomized clinical trials}

\label{erstes}
Evaluating statistical data from clinical trials 
is one of the most applied methods to investigate
the effect of drugs or certain therapies on the 
patient's health.
To compare the recovery rate of the patients that have taken the 
drug to the recovery rate of the others is among the most popular methods of
research in medicine.
Despite the simplicity of this method,
it can produce an abundance of misconclusions if it is not applied
carefully.
One of the most popular errors in too naive applications of this 
kind of statistical reasoning is not to distinguish clearly
between so-called {\it experimental} and {\it non-experimental} data.
Experimental data is produced if some (randomly selected) patients
are advised to take the drug and some are advised not to take it.
In the second case the patients may decide for their own whether they want
to take it or not. Although one may prefer the second method for
ethical reasons, the worth of the produced data for drawing causal
statements is considerably reduced. There is even no way to 
{\it prove} any causal effects of the drug by those data sets.
The reason is that the possibly observed correlation between 
recovery rate and  taking the drug may stem from a hidden common cause:
There may be an (unobserved) 
 feature of certain patients that makes them  decide
to take the drug and at the same time makes them recover. In the simplest 
case, this common
feature may be a variable that could be observable in principle.
Assume for instance, the elderly people tend less to take the drug than the
younger ones. On the other hand, they recover less likely.
This would produce a correlation that fakes a causal effect of
the drug.
But this misconclusion could be avoided by taking into account
the patient's age and evaluating the recovering rates separately
for each age. However, in the generic case, the common cause
is less simple and may even be something that is not accessible at all.
Imagine, for instance, there is something like a strong wish 
to recover without drugs that is highly correlated with a high
recovery rate. Then the variable describing the common cause
is a feature like personality (including mental and physiological 
constitution) and we are no able to quantify it in order to
compare only persons with ``equal personality''.
Randomized experiments  seem to avoid
this problem completely. However, there is still a problem
that is directly related to the problem above. 
 In typical tests we cannot expect that
all patients act like they are  advised to do: Some may 
not take it although they should and some may even acquire  it 
although they
were not advised  to take it.
Assume that it is even possible to prove (by blood-tests, 
for instance) that some patients did not comply. 
Unfortunately, even this does not solve the problem completely
since we cannot identify  which part of  
the correlations are caused by the common
personal feature influencing the patient's  decision and his 
recovery behavior
and which part of the correlation  is a causal effect of the  drug.
However, it is clear that the causal effect of the drug 
can {\it almost} be identified if {\it almost all}  patients comply since
this case approximates the perfectly randomized experiment.
This intuition shows that quantitative lower and upper bounds
on the causal effect of the drug can be given depending on 
the compliance rate. This has been investigated thoroughly 
in the literature (see \cite{Pearl:00} and references therein). 
First we rephrase
their  most intuitive conclusion. Imagine a naive researcher,
not aware of the non-compliance, observes the two binary variables
$Z$ and $Y$ where $Z=z_1$ or $Z=z_0$  means that the  patient had or 
had not been selected to take the  drug and $Y=y_1$ or $Y=y_0$ means that
the patient recovered or not, respectively.
Then he takes the difference between the probability to recover
if one was advised to take the drug minus the probability
to recover if one was not advised, i.e.,
\[
P(y_1|z_1)-P(y_1|z_0)
\]
as the (positive)  causal effect of the drug.
\footnote{Here we use the large sampling assumption, i.e.,
the sample size is large enough to estimate 
the joint distribution $P$ on all observed  random variables.
Issues of significance of correlations 
can therefore be neglected here.}
Now imagine that  a less naive statistician asks
for the maximal error that this naive 
conclusion can cause.
He observes the variable $X$ where $X=x_1$ or $X=x_0$ means that
the patient has or has not taken the drug, respectively.
He concludes that
in the worst case of overestimating the effect, 
all those that have been advised to take the drug
but  complied and have recovered, 
would have stayed ill if they had 
complied. On the other hand, all those that had been advised not to take
it and took it nevertheless and did not recover may have recovered
if they would have complied.
By this intuition he concludes that 
the causal effect of the drug
is to  
increase the probability to recover 
at least by
\begin{equation}\label{nat1}
P(y_1|z_1)-P(y_1|z_0)\,\,-P(y_1,x_0|z_1)-P(y_0,x_1|z_0)\,,
\end{equation}
where $P(y_1,x_0|z_1)$ denotes the conditional probability
of the event `no drug taken and recovered', given that the advice 
was to take it. The other definitions are similarly.

By  the same kind of arguments one can give bounds on the underestimation
of the causal effect. One obtains that the recovery rate is increased
by at most
\begin{equation}\label{nat2}
P(y_1|z_1)-P(y_1|z_0)\,\,-P(y_0,x_0|z_1)-P(y_1,x_1|z_0)\,.
\end{equation}
Note that the increase of recovery rate we are talking about
is the increase that would happen if
all patients took the drug, also those that have decided not to take it.
Therefore the definition of the causal effect  
relies on the hypothetical result of  an experiment where
all patients are forced to comply. 

For a formal proof of statements of this kind one needs a precise
model in which terms as 
``the recovery rate if all patients take the drug''
make sense. 

Here we follow the approach of Pearl \cite{Pearl:00} to formalize
causal claims of this kind within graphical models. 
Random variables $X_1,\dots,X_n$ 
are the nodes of directed acyclic graphs
and an arrow from variable $X$ to $Y$ indicates that there is a 
causal effect from $X$ to $Y$.
Furthermore, one  has to specify the transition probabilities for each node,
i.e. for each variable $X_j$ one has the conditional probabilities
\[
P(x_j|x_{j_1},\dots,x_{j_k})
\]
where $X_{j_1},\dots,X_{j_k}$ are the parents of $X_j$.
The joint distribution on all the random variables decomposes into
\[
P(x_1,\dots,x_n)=\prod_j P(x_j|x_{j_1},\dots,x_{j_k})\,.
\]

An essential part of Pearl's theory is that 
it distinguishes carefully between the probability measure
that is obtained once one has {\it observed} that the variable
$X_j$ takes the value $x_j$ and the measure that is obtained
if $X_j$ {\it is set} to $x_j$ by external control.
The first one is the usual conditional probability
\[
P(x_1,\dots ,x_n| x_j)\,,
\]
whereas he denotes the latter one by
\[
P(x_1,\dots,x_n| \hbox{ do } x_j)\,.
\]
He 
formalizes
this probability  by 
\[
P(x_1,\dots,x_n| \hbox{ do } \hat{x}_j):=
\prod_{i\neq j} P(x_i|x_{i_1},\dots,x_{i_k}) \delta(x_j,\hat{x}_j)\,.
\]
The intuition behind this definition is that the values of $X_j$ 
do no longer depend on the values of those variables that influence
$X_j$ directly (``the parents of $X_j$'') or indirectly
(``the ancestors of $X_j$''). Hence the transition probabilities
``probability that $X_j$ takes the value $x_j$, given the value of the 
parents of $X_j$'' have to be substituted by the Kronecker symbol.
This manipulation, for instance, 
does not change the probability distribution of the ancestors
of $X_j$.  It changes only the distribution of the descendants of $X_j$.
Whereas conditional probabilities reflect {\it correlations}
among variables, the probabilities obtained by the do-operator
reflect {\it causal} effects and distinguish between causal directions.

Following \cite{Pearl:00} 
the relevant variables in our drug testing problem
are $X,Y,Z$ as already introduced and a latent variable
$U$ that includes unobservable
 features as personality and physical constitution.
The graphical model of the causal structure is shown in Fig.~1.

\begin{figure}
\centerline{
\epsfbox[0 0 110 120]{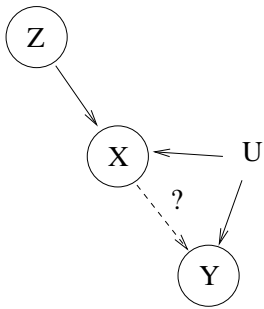}
}
\caption{\small Graphical model of causal structure of the compliance problem:
$Z$ is the advice to take the drug, $X$ is patient's 
decision to take it, $Y$ is his physical response (recovered or not), 
and $U$ represents all relevant latent factors influencing
decision and response.} 
\end{figure}

Whether there is an arrow from $X$ to $Y$ should be clarified 
by the clinical trial. A priori, we assume that there is an arrow.
Then the model is specified by the following parameters 
\[
P(z),\,\, P(x|z,u),\,\, P(y|x,u),\,\, P(u)\,.
\]
Whether or not and how $P(y|x,u)$  dependends really on $x$
is not clear yet.
The calculation of $P(y| \hbox{ do } x)$ 
would require to know these parameters, whereas 
only the joint distribution on $X,Y,Z$ is observable.
But  one can find bounds on
the so-called average causal effect  \cite{Pearl:00}
\begin{equation}\label{ACEDef}
ACE(X\rightarrow Y):= 
P(y_1| \hbox{ do } x_1) - P(y_1| \hbox{ do } x_0 )=
\sum_u \Big( P(y_1|x_1,u)-P(y_1|x_0,u)\Big) P(u)
\end{equation}
in terms of observable quantities.
Above, we have already mentioned one upper and one lower bound 
rather informally, namely inequalities  (\ref{nat1}) and (\ref{nat2}).
Following \cite{Pearl:00} we will call them  
 ``natural bounds''. They can formally be proven by the following
considerations.
As argued in \cite{Pearl:94}  one can chose $U$  w.l.o.g. such that 
it can attain $16=4\times 4$ possible values and such that 
$U$ determines $X$ and $Y$ deterministically if the actual value $z$ of $Z$ 
is given
(compare also \cite{BaPearl:94a,BaPearl:94b}).
The values of $U$ can be considered as a cartesian product
of a variable $B$  that determines the patient's 
decision $X$ and a component $R$  that determines
$Y$, i.e., the physical response to the drug,i.e., to recover of not.
The values of the first component 
are ``never take'',  ``always take'', ``comply'' and 
``non-comply'' \cite{ImRub:97}. 
Following \cite{HeckShach}
the response behavior to the drug is given by 
``never recover'', ``always recover'', ``helped'', and ``hurt''.
We denote these 4 values by $R=r_1, R=r_2, R=r_3, R=r_4$.
In this deterministic model, the only free parameters
remaining are $P(z)$ and $P(u)$. 
Then the average causal effect is given by 
the difference \cite{Pearl:00} 
\[
P(r_3) -P(r_4)\,,
\]
i.e., the probability to {\it recover  
because of} the drug minus the probability
{\it not to recover  because of} the drug.
Counterfactual statements like
``the patient would have recovered without using the drug''
are subject of philosophical debates with a long history 
(see \cite{Pearl:00}).
Therefore it may cause unease to some readers to
find causal implications of data based on such terms.
However, Pearl has come up with convincing arguments (to our opinion)
that counterfactual statements do make sense if appropriate
models are available.
Provided that one accepts that all variables describing
mental and physiological processes are classical random variables
the conclusion that it can be represented w.l.o.g. by a $16$-state variable
seems indubitable.
 However, the assumption that all mental and physiological states  
(influencing human decision and health)
are described by   
classical variables, is not self-evident at all. Here we do not 
participate in the (sometimes speculative) debate on the
relevance of quantum superposition and incompatibility of quantum observables
for mental processes \cite{Penr,KleinS}.
We just want to find out which rules in classical statistics 
can be violated {\it in case} quantum uncertainty is relevant for mental and 
physiological processes.

\section{The instrumental inequalities}

Within the hidden variable model 
where $U$ can attain $16$ different values 
one can prove the natural bounds (\ref{nat1}) and (\ref{nat2})
and even tighter bounds \cite{Pearl:00}. 
These so-called instrumental inequalities read \cite{Pearl:00}:
\[
ACE \geq \left\{\begin{array}{c} P(y_1,x_1|z_1) + P(y_0,x_0|z_0) -1 \\
P(y_1,x_1|z_0) + P(y_0,x_0|z_1) -1 
\\
P(y_1,x_1|z_0) -P(y_1,x_1|z_1)-P(y_1,x_0|z_1)-P(y_0,x_1|z_0)-P(y_1,x_0|z_0)\\
P(y_1,x_1|z_1) -P(y_1,x_1|z_0)-P(y_1,x_0|z_0)-P(y_0,x_1|z_1)-P(y_1,x_0|z_1)\\
-P(y_0,x_1|z_1)-P(y_1,x_0|z_1)\\
-P(y_0,x_1|z_0)-P(y_1,x_0|z_0)\\
P(y_0,x_0|z_1)-P(y_0,x_1|z_1)-P(y_1,x_0|z_1)-P(y_0,x_1|z_0)-P(y_0,x_0|z_0)\\
P(y_0,x_0|z_0)-P(y_0,x_1|z_0)-P(y_1,x_0|z_0)-P(y_0,x_1|z_1)-P(y_0,x_0|z_1)
\end{array}\right\}
\]

Note that the  problem to identify the causal effect
\[
P(y_1| \hbox{ do } x_1) - P(y_1| \hbox{ do } x_0)=
P(y_0| \hbox{ do } x_0) - P(y_0| \hbox{ do } x_1)  
\]
is symmetric with respect to  the joint substitution
$x_1 \leftrightarrow x_0$ and $y_1\leftrightarrow y_0$.
Furthermore it is symmetric with respect to the substitution
$z_1\leftrightarrow z_0$ since, abstractly considered, 
$Z$  is only an arbitrary
binary variable that influences the patient's decision. Note that perfect 
non-compliance would also make it possible to identify the effect 
of the drug. The 8 inequalities form
two groups, namely $1,2,5,6$ and $3,4,7,8$ such that members 
of the same group can be converted into each other by those symmetries.

The upper bounds on the causal effect are given similarly 
by the substitution $y_1 \leftrightarrow y_0$ and reversing the signs
of all probabilities:

\[
ACE \leq \left\{\begin{array}{c} 1- P(y_0,x_1|z_1) - P(y_1,x_0|z_0) \\
1- P(y_0,x_1|z_0) - P(y_1,x_0|z_1) \\
-P(y_0,x_1|z_0) +P(y_0,x_1|z_1)+P(y_0,x_0|z_1)+P(y_1,x_1|z_0)-P(y_0,x_0|z_0)\\
-P(y_0,x_1|z_1) +P(y_1,x_1|z_1)+P(y_0,x_0|z_1)+P(y_0,x_1|z_0)-P(y_0,x_0|z_0)\\
P(y_1,x_1|z_1)+P(y_0,x_0|z_1)\\
P(y_1,x_1|z_0)+P(y_0,x_0|z_0)\\
-P(y_1,x_0|z_1)+P(y_1,x_1|z_1)+P(y_0,x_0|z_1)+P(y_1,x_1|z_0)+P(y_1,x_0|z_0)\\
-P(y_1,x_0|z_0)+P(y_1,x_1|z_0)+P(y_0,x_0|z_0)+P(y_1,x_1|z_1)+P(y_1,x_0|z_1)
\end{array}\right\}
\]
These bounds imply the natural bounds (\ref{nat1}) and (\ref{nat2})
(for details see \cite{Pearl:00} and references therein).

\section{Quantum model of latent factors}

\label{Qmodel}

Here we do not want to discuss the difficult question to what
extent quantum mechanical effects play a crucial role for mental and 
physiological processes. However, we do not want to base causal
conclusions from statistical data on the assumption, that
quantum effects are irrelevant in our brain and our body.
For doing so, we have to propose a model that generalizes the
model above from classical to quantum probabilities. 
We chose
the formal setting of algebraic quantum theory which is general enough
to include quantum and classical physical processes.

In this setting,
the observable algebra of an arbitrary physical system is described
by a $C^*$-algebra   $\cA$  (see \cite{BR1,Mu90})
containing the unity  ${\bf 1}$. The algebra 
$\cA$ is called  the ``algebra of observables''.
In a pure quantum system with unrestricted superposition principle
(``without super-selection rules'')
 $\cA$ may for instance be the algebra of bounded linear operators
on an arbitrary Hilbert space.
An element $a\in \cA$ is positive, written $a\geq 0$ 
 if it can be written as
$a=b b^*$. A functional 
\[
\rho :\cA \rightarrow \C
\]
 is called {\it positive} 
if it maps positive elements on non-negative numbers. 
The {\it states} are positive functionals  $\rho$ of
norm 1, where the norm is described by 
\[
\|\rho\|:=\sup_{a\in \cA}| \rho(a) |/\|a\| =|\rho({\bf 1})|\,,
\]
and $\|a\|$ denotes the operator norm of $a$.
Every {\it yes-no experiment}, i.e. an experiment with two  
possible outcomes, is described by a positive operator $a$ with $a\leq 1$,
i.e., $1-a\geq 0$ and $\rho(a)$ is the probability for the outcome ``yes''
if the system is in the state $\rho$.

A physical process changing the system's state
 can either be described by a completely-positive map $G^*$ (``CP-map'') 
\cite{Da76} on the set of positive functionals
with $\|G^*(\rho)\|=\|\rho\|$
or, by duality, as  CP-map 
\[
G:\cA\rightarrow \cA
\]
with $G({\bf 1})={\bf 1}$.
In this formulation, the process
acts on the state by transforming $\rho$ to $G^*(\rho)\rho\circ G$.
Measurements with any finite number $k$ of  outcomes 
are described by so-called positive operator valued 
measures (``POVMs''), i.e.,
a family $m_1,m_1,\dots,m_k$ of positive operators with $\sum_i m_i={\bf 1}$.
Then
$\rho(m_i)$ denotes the probability to obtain the result ``$i$''.
Note that the POVM does not describe the effect of the measurement 
instrument on the state. If the state after the measurement is relevant, we
have to describe the instrument by a family of
$k$ CP-maps $G_1,\dots,G_k$ with $\sum_i G_i({\bf 1})={\bf 1}$.
If the measurement outcome is ``$i$'' the state $\rho$ is transformed to
$\tilde{\rho}$ with
\[
\tilde{\rho}(a):=\rho( G_i (a))/\rho(G_i({\bf 1}))\,,
\]
where
$\rho(G_i({\bf 1})$ is the probability to obtain the result
``$i$''. Hence 
\[
G_1({\bf 1}),\dots,G_k({\bf 1})
\]
is the 
POVM representing the measurement.
Without observing the   measurement outcome,
the instrument  transforms $\rho$ to $\rho \circ G$ with $G:=\sum_i G_i$.

The formal setting for investigating the violation
of the instrumental inequalities by quantum latent factors
is based on the following assumptions.

\begin{enumerate}

\item
The advice to take or not to take the drug is perfectly randomized
and independent of all other factors.

\item All relevant latent factors influencing the patient's
decision and his response behavior to the drug
are described by the state of a physical system in the sense
of algebraic quantum theory.
This state includes the patient's mental and physical state
as well as noise that influences the decision or
response or both.
The state is the
state $\rho$  of a physical system (in the sense above)
described by an observable algebra $\cA$.
The system  
is either purely 
quantum, purely classical, or a mixture of both. 
Although this may be a too materialistic view on mind and consciousness
this approach is more general than any hidden variable model
in the literature.

\item
To take or not to take 
the drug is a classical event that either 
occurs or does not occur but there is no
quantum superposition between both. The process of human decision
is therefore like a measurement  process in its broadest sense
explained above. This instrument is described by CP-maps  $D_1,D_0$
acting on $\cA$. 
Hence the state $\rho$ is transformed to 
$\rho \circ D_1/\|\rho\circ D_1\|$ if the patient
has decided to take the drug and $\rho\circ D_0/\|\rho\circ D_0\|$ 
otherwise.
If the decision itself is ignored, the process of decision making
is described by the process  $\rho\mapsto \rho \circ D$ with
$D:=D_0 +D_1$.

\item
The advice to take or not to take the drug is a classical signal
that influences the patient's internal state. 
The advice to take or not to take transforms the  state to $\rho\circ G_1$
or $\rho\circ G_0$, respectively. Here $G_j$ are CP-maps on $\cA$.

\item The effect of the drug is to transform the internal state
$\rho$ to $\rho\circ E_1$, whereas the natural evolution without drug 
in the considered
time interval
changes the state according to the operation 
$E_0$. The operations $E_j$ are CP-maps on $\cA$.

\item
Whether the patient recovers or not is a classical event and is therefore
equivalent to a measurement process in the sense above.
It corresponds to a yes-no-experiment described by a positive operator 
$m \in \cA$.

\item 
The advice to take or not to take the drug has no 
direct causal influence on the heath, it influences the
probability of
recovery only indirectly by influencing the decision.
This corresponds to the fact
that  the graphical model Fig.~1 for the classical setting
has no arrow from $Z$ to $Y$.

\end{enumerate}

One may think that it would be more appropriate  to
assume that the operations $G_j$ and $E_j$ act on different systems:
$G_j$ acts on the mind and $E_j$ on the body. But we do not want to restrict
our proofs to this assumption. In particular, we emphasize that 
there may be a part of the body with the property that its quantum state
influences the decision and the recovery. This may, for instance, be
a cell that influences  the production of some hormone that
has a causal effect on both mood  and health.

For the observable quantities we obtain:
\begin{equation}\label{Gbeob1}
P(y_1,x_k|z_j)= \rho (G_j D_k E_k (m)) 
\end{equation}
is for $k=1$ ($k=0$) 
the probability to take (take not) the drug 
and recover, given that the advice was to take the drug, i.e. $j=1$ 
(not to take, $j=0$).
Similarly we have
\begin{equation}\label{Gbeob2}
P(y_0,x_k|z_j)= \rho (G_j D_k E_k ({\bf 1} -m)) \,.
\end{equation}
For causal statements the following 
unobservable counterfactual  probabilities are
important:
\[
\rho(G_j D_k E_l (m))
\]
with $k\neq l$. It expresses the hypothetical experiment that
we observe the patients decision to take or not to take the drug
and prevent him from taking it although he has decided to
or force him to take it although he has refused to.
In the following we use the abbreviation $m_l:=E_l(m)$.
Assumption 7 translates to the statement
\begin{equation}\label{Irre}
\rho(G_1 D (m_l))=\rho(G_0 D (m_l)) \,,
\end{equation}
for $l=0,1$.

The increase of the recovery probability that is caused 
by the drug (ACE) is given by
\[
ACE=\rho(G_1 D(m_1))- \rho(G_1 D(m_0))=\rho(G_0 D(m_1))-\rho(G_0 D(m_0))\,.
\]
Note that 
$\rho(G_1 D(m_1))$ is a sum of the observed probability
$\rho(G_1 D_1(m_1))$ and the counterfactual probability 
$\rho(G_1 D_0(m_1))$. This reflects the fact that
the causal effect of the drug could only be identified  
if the taking of the drug was 
decoupled from 
the patient's decision to take it.
In the next section we will present a
model violating the third instrumental inequality. Due to the symmetry 
of the problem we can violate inequalities $4,7,8$ similarly.

\section{Violation of instrumental inequalities 
and Bell inequality}

The violation of the so-called 
Bell inequality  \cite{Bell} is one of the most  convincing
arguments supporting the hypothesis that micro-physics 
cannot be described by classical probability theory.
The idea behind Bell's inequality is that it describes
quantitatively the difference between those statistical 
correlations that appear in quantum theory and those that
can be explained by a classical probability space where 
all uncertainty stems from our missing knowledge on 
the values of  some hidden parameters.
These parameters 
should decide deterministically which event will occur in future. 
Whereas Einstein, Podolsky, and Rosen 
(in the so-called EPR-paradox)
gave intuitive arguments
why quantum correlations may behave rather strange,
Bell's inequality  formalized a testable difference between
quantum and classical correlations. 

As already noted in \cite{Pearl:00}
the instrumental inequalities have some formal analogies 
to Bell's inequality
since they give bounds on the possible 
correlations between two random variables
that are influenced by a common hidden parameter.
We took this ``formal analogy'' seriously and show that the well-known 
setting that shows the violation of Bell's inequality
can be modified to violate the instrumental inequalities 
$3,4,7,$ and $8$.

We consider 
a quantum system in
$\C^4$ and decompose it into $\C^4=\C^2\otimes C^2$. The two
basis states of each subsystem can be for instance  the
horizontal and vertical polarization of a photon, denoted by
$|h\rangle$ and $|v\rangle$, respectively.
The superposition
\[
\cos (\theta) |h\rangle + \sin (\theta) |v\rangle
\]
describes a polarization in the direction of an axis 
with the angle $\theta$  
with respect to the horizontal axis.
The polarization can be measured by a polarization filter.
All photons that pass the filter
are polarized in the direction of the axis of the filter.
A photon with polarization $\theta$ has the probability
$1/2(1+\cos( 2\theta-2\tilde{\theta}))$ 
to pass the filter  if $\tilde{\theta}$ is the
direction of the filter's axis.
We write the measurement result ``$1$'' if it passes and ``$0$'' 
if it doesn't.
Then we consider the so-called singlet state
\[
|\psi\rangle:= \frac{1}{\sqrt{2}}(|h\rangle|v\rangle-|v\rangle |h\rangle)   \,.
\]
It has the following interesting property:
For each filter both results appear with equal probability.
If the  polarization axis of filter 1 and filter 2  have the angles 
$\alpha$ and
$\beta$, respectively, the probability that both measurement results
coincide is 
\begin{equation}\label{Corr}
\frac{1}{2}(1- \cos (2\alpha-2\beta))\,.
\end{equation}
For $\alpha=\beta$ this means that the results are always different
and for $\alpha=\beta \pm 90^o$ they always coincide.

In the typical setting to show that those kind of quantum correlations
are fundamentally
different from classical correlations there is a source emitting a photon pair
in a singlet state in two different directions and the polarization
axis of the two filters are chosen randomly and independently.
Bell has shown \cite{Bell} 
that there exist angles $\alpha_0,\alpha_1$ for the first 
polarization analyzer and $\beta_0,\beta_1$ for the second such that
the correlations (\ref{Corr}) cannot be explained by
any hidden variable theory that is local. Here locality means that 
causal effects between physical systems require physical signals
traveling through the space not faster than the speed of light.
Explicitly, Bell's inequality is as follows.
Assign the values $e_j=\pm 1$ to the result ``photon has passed the 
filter $j$'' or ``photon has not passed'', respectively.
Then define the covariance 
\begin{equation}\label{Cova}
C(\alpha,\beta):= \sum_{e_1=\pm 1,\, e_2=\pm 1} e_1 e_1 P(e_1,e_2|\alpha,\beta)\,,
\end{equation}
where $P(.|\alpha,\beta)$ describes the joint distribution on
the measurement outcomes given the position of the filters.
Then Bell's inequality states that
\[
|C(\alpha_0, \beta_0)+ C(\alpha_0,\beta_1)+ C(\alpha_1,\beta_0)- 
C(\alpha_1,\beta_1)| \leq 2
\]
is satisfied whenever any  hidden variable determines both measurement 
results 
in advance, i.e.,
before one  has chosen the filter angles. 
In contrast, with the singlet state one can achieve
the values $\pm 2\sqrt{2}$
if the angles are chosen appropriately.
The factor $\sqrt{2}$ describing the difference between quantum 
and classical correlations will also appear in 
the violation of instrumental inequalities below.

Meanwhile,
experiments of this kind 
\cite{Aspect:81,Gisin:98} have  given strong evidence for the fact
that micro-physical effects 
are really in good agreement with quantum theory
and
hence in contradiction to any local
hidden variable theory.
Since  the experiment above  is one of those that 
cannot be explained by any local hidden 
variable model it is straightforward to use this setting to 
construct a causal toy  model 
for patient's  decision and his physical response that violates
the instrumental inequalities.
Let
the left polarization filter initially have the angle  $\alpha_0$. 
The advice to take the drug turns the filter
to the angle $\alpha_1$.
The result of the left polarization measurement is the patient's decision.
The right filter is initially in the position $\beta_0$. The drug
turns the filter to the position $\beta_1$.
The measurement of the right photon decides whether the patient recovers.
This is shown in Fig.~2

\begin{figure}
\centerline{
\epsfbox[0 0 424 285]{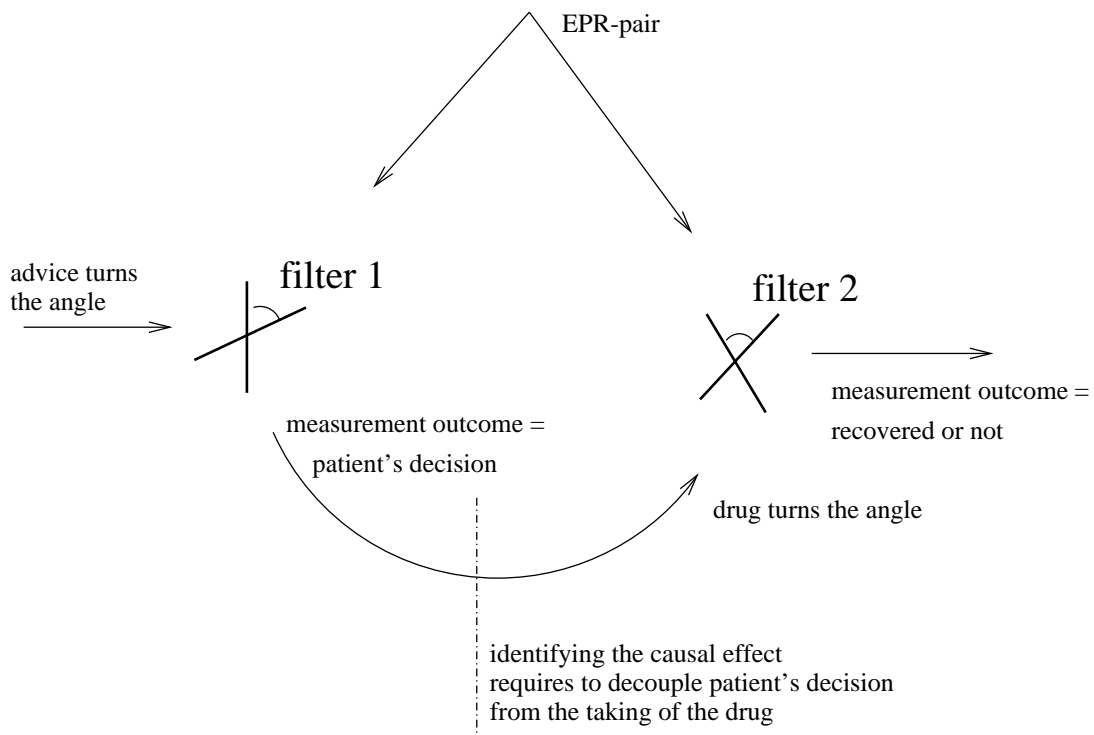}
}
\caption{\small Unrealistic toy model of  human behavior. The 
process of decision making and the process determining the 
physical  response to the drug
are influenced by a common quantum state.}
\end{figure}

In this setting the  decisive conditional probabilities are given by
\[
P(y_j,x_k|z_l)=\frac{1}{4}\Big( 
1- (-1)^{j+k} \cos( 2 \alpha_l - 2\beta_k)\Big)\,.
\]
This can be seen as follows:
The indices $l$ (advice to take/ not to take)  
and $k$ (taken or not)  determine the angles $\alpha$ and $\beta$
of the filters. The probability that 
the patient's decision ($0$ or $1$)
coincides with his response to the drug ($0$ or $1$)  
is given by 
$( 1- \cos( 2 \alpha_l - 2\beta_k)/2$. The probabilitiy that 
the results  disagree is 
$( 1+ \cos( 2 \alpha_l - 2\beta_k)/2$. The probabilities for the results
$1$ and $0$ in the first measurement are $1/2$ each, regardless of the
measurement angle. This gives an additional factor $1/2$.
Here we will not really  need the general setting using CP-maps 
as described in Section \ref{Qmodel}. 
In order to show, that the experiment
described above fits in the general formalism we shortly give the
definitions of the CP-maps $G_j,D_k,E_l$ in an  informal way.
The maps $G_0$ and $G_1$ describe the 
turning of the first filter by the angle  $\alpha_0$ or $\alpha_1$, 
respectively. 
$D_0$ and $D_1$ describe the operations on the system caused
by a vertical polarization measurement if the result is 
negative or positive, respectively. 
$E_0$ and $E_1$ correspond
to the filter rotations by the angles $\beta_0$ and $\beta_1$, respectively.
The yes-no experiment that is given by polarization a measurement
of the second photon (with respect to the vertical axis) is described
by the operator $m$. Hence the operators $E_0(m)$ and $E_1(m)$ describe
polarization measurements with angles $\beta_0$ and $\beta_1$, respectively.
Let $P^\alpha_1$ and be the projector 
onto the subspace of $\C^2$ spanned by 
\[
\cos \alpha |h\rangle
+\sin \alpha |v\rangle
\]
and $P^\alpha_0$ be the projector
onto the orthogonal subspace.
Then $G_j \circ D_k$  is the CP-map
\[
a \mapsto (P^{\alpha_j}_k \otimes {\bf 1}) \, a \,  
(P^{\alpha_j}_k \otimes {\bf 1})\,.
\]
Here we have assumed for simplicity that the polarizator 
is not a filter that absorbs some photons and let the others pass
but we assume that is {\it splits} the photon beam into 
those with vertical and those with horizontal polarization. 
Otherwise the operation
$D_0$ would be more difficult to describe formally since it annihilates the 
photon completely.
The positive operators $m_j=E_j(m)$ are given by
\[
E_l (m) ={\bf 1} \otimes P^{\beta_l}_1 \,.
\]
By applying the CP maps $G_j$ and $D_j$ to $m_l$ we obtain 
\[
G_j D_k (m_l) = P_k^{\alpha_j} \otimes P_1^{\beta_l}\,.
\]
Due to 
\[
P^{\alpha_j}_1+P^{\alpha_j}_0={\bf 1}
\]
we have 
\[
G_0D (m_l)= G_1 D (m_l) = {\bf 1} \otimes P_1^{\beta_l}\,.
\]
This shows that equation (\ref{Irre}) is satisfied.

It is almost obvious, that the average effect of the drug
is zero since in both filter positions the
probability to pass is $1/2$.
If we would decouple the taking from the patient's decision
the recovery probability was in both cases $1/2$  whether  the drug was 
 taken or not.

This does not mean that the drug does not have any causal effect on 
the patient's health at all: 
The drug might help the patients that have decided 
to take it  and could hurt the others if they had been
forced to take it.
Hence  
the effect is only zero in the average for a hypothetical experiment
where all patients are forced to take the drug, even those that 
would have decided against it.
Since we have already argued that the average causal effect is zero,
the third instrumental inequality is violated if we find polarizator
positions $\alpha,\beta,\gamma, \delta$ such that 
\begin{equation} \label{Ins3}
P(y_1,x_1|z_0) -P(y_1,x_1|z_1)-P(y_1,x_0|z_1)-P(y_0,x_1|z_0)-P(y_1,x_0|z_0)
>0 \,.
\end{equation}
For doing so, we show that we can chose the angles such that
\[
P(y_1,x_1|z_0)= 1/4(1+1/\sqrt{2})=:a^+
\]
and all the other four terms should be
\[
1/4(1-1/\sqrt{2})=:a^-\,.
\]
The equation $P(y_1,x_1|z_0)=a^+$  is satisfied 
if 
\[
\alpha_0 -\beta_1=90^o\pm 22,5^o\,.
\]
By setting 
\[
\alpha_1-\beta_1=\pm 22,5^o
\]
we achieve that
$P(y_1,x_1|z_1)=a^-$. With 
\[
\alpha_1-\beta_0=90^o\pm 22,5^o
\]
we have $P(y_1,x_0|z_1)=a^-$. By  
\[
\alpha_0-\beta_1 =90^o\pm 22,5^o
\] 
we obtain
$P(y_0,x_1|z_0)=a^-$. By 
\[
\alpha_0-\beta_0=90^o\pm 22,5^o
\]
 we achieve
$P(y_1,x_0|z_0)=a^-$.
All these equations can be 
satisfied if 
\[
\beta_1=0^o ,\,\,\,\alpha_0=67,5^o ,\,\,\,
\beta_0=-45^o,\,\,\, \alpha_1=22,5^o\,.
\]
Then the left hand side of inequality (\ref{Ins3}) is
\[
a^+-4a^-= 1/4(-3+5/\sqrt{2}) \approx 0.134 \,.
\]
Hence the third instrumental inequality  claims the average
causal effect to be about $13\%$ although it is zero.
Note that in this setting it was essential that 
there is indeed a causal effect of the drug on the recovery -
sometimes negative and sometime positive depending on the patient's 
decision.
Hence the conclusion that the drug influences patient's health 
is true nevertheless.

However, we present a modified version of the Gedankenexperiment
where every classical statistician should come to the conclusion that
there is a causal effect although the drug does not influence health
at all. This version is even more analogue to the violation
of Bell's inequality, actually it is just a reinterpretation
of it. 
Assume as above that some randomly selected patients are advised to take
or not to take a drug, respectively. Not everyone complies
but we can identify the non-compliers afterwards.
After waiting for a while
some of the patients (randomly selected) 
are given another drug. Since it is given in the presence
of the doctor, we exclude the possibility of noncompliance
here. Then we observe which patients have recovered. 
We describe 
the experiment by $4$ observed binary random variables $X,Y,Z,W$,
where $X,Y,Z$ are as in Section \ref{erstes}
and $W$ is the taking of the second drug.
As in Section \ref{erstes} the
there is a latent variable $U$ that influences
decision and recovery behavior.
The complete causal structure is given by the 
graphical model in Fig.~3.

\begin{figure}
\centerline{
\epsfbox[0 0 250 130]{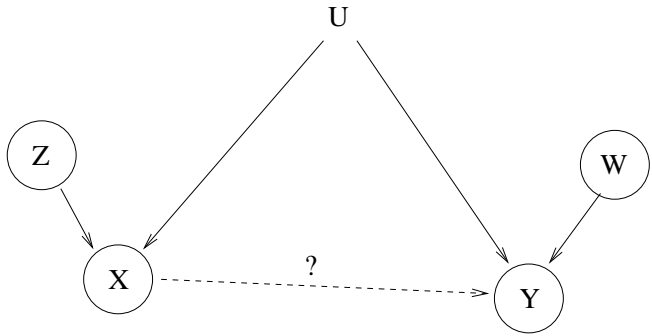}
}
\caption{\small Graphical model of causal structure between
the advice to take the first drug ($Z$), the taking of it ($X$),
the taking of the second drug ($W$), and the recovery ($Y$).
The hidden variable $U$ represents all relevant latent factors
influencing decision and recovery.}
\end{figure}

Whether or not there is an arrow from $X$ to $Y$ should be clarified
by statistical  data.

Now we consider once more the singlet states and
assume that  the advice to take the drug turns the filter position
from $\alpha_0$ to $\alpha_1$. As above, the result of the measurement 
at the left filter is the patient's decision. In contrast to the
setting above, the taking of the drug does not have any effect on the 
second filter. 
The
second filter is turned from angle $\beta_0$ to $\beta_1$ 
by the second drug.
With the definition of eq. (\ref{Cova})
the singlet state can produce a joint measure on
the outcomes such that
\[
C(\alpha_0,\beta_0) +C(\alpha_1,\beta_0)
+C(\alpha_0,\beta_1)-C(\alpha_1,\beta_0)= \pm 2\sqrt{2}\,,
\]
for appropriate polarizator angles $\alpha_j$ and $\beta_j$.
In analogy to Bell's original argument, this value
for the sum of covariances  cannot be explained by
any classical variable $U$ in the graphical causal model of Fig.~3
if
there is no causal effect
from $X$ to $Y$.
 The classical statistician
should therefore  come to the erroneous 
conclusion that there must be a causal
effect from the first drug on the recovery.

Admittedly, we do not know of any example, where 
classical statistics draw
causal conclusions based on
Bell's inequality. However, we want to point out that causal
conclusions based on classical probability theory
may even fail if the physiological and mental
processes that are decisive for human behavior are 
sensitive to {\it quantum} noise. It is not necessary
that the mental and physiological processes themselves
are non-classical.

\section{Some instrumental inequalities are still valid}

We will prove that the  group $1,2,5,6$ of instrumental inequalities 
is still valid in the quantum setting of Section \ref{Qmodel}.
Due to the symmetry of the problem it is sufficient to prove only
the first one
 given by
\[
ACE \geq P(y_1,x_1|z_1) + P(y_0,x_0|z_0) -1\,.
\]
It can be  shown using simple operator inequalities.
We have to show that
\begin{equation}\label{zuzeigen}
\rho(G_1D(m_1))-\rho(G_1D(m_0))\geq 
\rho(G_1 D_1 (m_1)) +\rho(G_0 D_0 ({\bf 1} -m_0)) - 1\,.
\end{equation}
We show this by simple calculations with operators.
Note that
\begin{equation}\label{start}
G_0 D_1 ({\bf 1}-m_0) \geq 0\,,
\end{equation}
since ${\bf 1}-m_0$ is a positive operator. The reason is that
${\bf 1}-m$ is positive  and ${\bf 1} -m_0={\bf 1}-E_0(m)= E_0({\bf 1} -m)$
is positive because it is obtained by the application of the
CP-map $E_0$ on ${\bf 1}-m$.
Similarly, the application of the CP-maps $D_1$ and $G_0$ conserve
positivity. 
By the same arguments, $G_1 D_0(m_1)$ is positive. Hence we obtain
\begin{equation}\label{start2}
G_0 D_1 ({\bf 1}-m_0) + G_1 D_0(m_1)  \geq 0\,.
\end{equation}

Inequality (\ref{start2}) is equivalent to
\begin{equation*}
G_0 D_0 (m_0) + G_0 D_0({\bf 1}-m_0) + G_0 D_1 ({\bf 1}-m_0) +G_1 D_0(m_1) 
\geq G_0D_0 ({\bf 1})
\end{equation*}
Using $D_0+D_1=D$ we obtain
\[
G_0 D_0 (m_0) + G_0 D({\bf 1}-m_0) +G_1 D_0(m_1) \geq G_0D_0 ({\bf 1})\,.
\]
This is equivalent to
\[
G_0 D_0 (m_0) +G_1 D ({\bf 1})+G_1 D_0(m_1) 
\geq G_0D_0({\bf 1})
+G_1D (m_0) \,.
\]
By $G_1D({\bf 1})={\bf 1}$  we obtain
\[
G_0 D_0 (m_0) +{\bf 1} +G_1 D_0(m_1) +G_1D_1(m_1)\geq 
G_0 D_0 ({\bf 1}) + G_1 D (m_0)+G_1D_1(m_1)\,,
\]
and
\[
G_1 D (m_1) - G_1 D (m_0) \geq
 G_1 D_1 (m_1) + G_0 D_0 ({\bf 1}-m_0) -{\bf 1}\,.
\]
Applying the state $\rho$ to both sides, we obtain inequality
(\ref{zuzeigen}).
The second, fifth, and sixth instrumental inequality follow
similarly due to the two symmetries of the problem according 
to the substitution
$z_1\leftrightarrow z_0$ 
(corresponding to the substitution $G_1\leftrightarrow G_0$)
and a common substitution $ x_1\leftrightarrow 
x_0,
y_1\leftrightarrow y_0$.
Note that the substitution $y_1\leftrightarrow y_0$ corresponds to the 
substitution $m_j\leftrightarrow {\bf 1} - m_j$ and
$x_0\leftrightarrow x_1$ to the substitutions 
$D_0\leftrightarrow D_1$ and $m_1 \leftrightarrow m_0$.
Hence the second symmetry corresponds to the operator
substitutions $D_1\leftrightarrow D_0$ 
and $m_1 \leftrightarrow 1-m_0$.

The same techniques can be used to prove the more intuitive natural
bound (\ref{nat1})
\begin{equation}\label{nat3}
ACE \geq P(y_1|z_1)-P(y_1|z_0)\,\,-P(y_1,x_0|z_1)-P(y_0,x_1|z_0\,.
\end{equation}
Note that the probability $P(y_1|z_1)$ is given by
$\rho(G_1D_1(m_1) +G_1D_0(m_0))$ and $P(y_1|z_0)$ is given by
$\rho(G_0D_1(m_1) +G_0D_0(m_0))$.
The probabilities that appear as the 
third and the fourth term in (\ref{nat3}) 
can be obtained from eq. (\ref{Gbeob1}) and eq. (\ref{Gbeob2}).

We show that the operator inequality corresponding to
inequality (\ref{nat1}) is true which reads
\begin{eqnarray*}
&&G_1D(m_1) -G_1 D(m_0) \geq \\&&G_1 D_1 (m_1) + G_1D_0(m_0)
-G_0 D_0 (m_0) -G_0 D_1 (m_1) -G_1 D_0 (m_0) -G_0 D_1 ({\bf 1} -m_1)\,.
\end{eqnarray*}
Due to  $G_1 D(m_0)=G_0 D(m_0)$ this is equivalent 
(by some elementary calculations) to
\[
G_1 D_0 (m_1) +G_0 D_1 ({\bf 1}-m_0)  \geq   0\,,
\]
which is certainly true.
Hence the most intuitive bound, which is probably the best  known
one, cannot be violated by quantum effects. This
is another interesting result.

\section{No large-scale entanglement required}

The EPR-setting above suggested that the violation of the instrumental 
inequalities requires large-scale entanglement between
the part of the brain where the decision is made
and the part of the 
body where the disease is located. 
We emphasize that large-scale entanglement of this kind is 
not necessary. Consider two nervous cells, one sending a signal to
any endocrine gland 
that produces a hormone that supports the recovery process
and the other one that sends signals to the central nervous system
and influence human decisions.
We assume that the output of both cells is a classical signal
but its internal state cannot completely described by classical
variables. Hence the internal process that decides 
whether the cell sends a signal or not is a quantum measurement process.
Then our setting requires that both cells share an entangled
quantum state and that their input signals operate on 
the corresponding internal quantum system.
Our considerations 
show that entanglement between two cells
may even produce interesting effects.
Such an entangled state may for instance be
provided by quantum correlated noise 
influencing both cells. Note that 
the EPR-pair provided by the
noise may even
be ``one-particle-entanglement'', i.e., a particle being in a superposition
of ariving at one cell and ariving at the other.

Nevertheless we do not want to speculate whether such models 
may be realistic.
It is not really plausible that quantum noise influencing
a small number of cells would produce correlations 
like in our toy model. One may also object that
 there is a large period of time between the decision
and the response to the drug. Hence the violation of 
the instrumental inequalities requires quantum coherence
that is stable for a long time (compared to time scales 
of  decoherence  in technical quantum systems).
Therefore the violation seems to be even less likely.

However,
the main purpose of this paper was to show that
some 
classical causal interpretations of  every-day life statistical data could
{\it in principle} fail
if latent quantum effects influence our behavior.

\vspace{0.3cm}

Thanks to Pawel Wocjan for useful corrections.

\end{document}